\documentclass[11pt]{article}
\usepackage{geometry}                
\usepackage{graphicx}
\usepackage{amssymb}
\usepackage{epstopdf}
\usepackage{latexsym}
\usepackage{amsmath}

\title{Quantum Field Theory in de 
Sitter space : A survey of recent approaches.}
\author{Jean Pierre Gazeau\thanks{gazeau@apc.univ-paris7.fr} and M. Lachi\`eze Rey\thanks{marclr@cea.fr}\\
  \emph{Astroparticules et Cosmologie\thanks{``APC'', UMR 7164 (CNRS,Universit\'e Paris 7, CEA, Observatoire de Paris)}, Boite 7020}\\
\emph{Universit\'e Paris 7 Denis Diderot, 2 Place Jussieu, 75251 Paris 
Cedex 05 Fr.}}

\def\bb{\mathbb}

\font\mybbi=msbm10 at 9pt
\def\bbi#1{\hbox{\mybbi#1}}
\def\barra#1{\not \!#1}

\def\NU{\Upsilon}
\def\Z{\bb Z}
\def\HH{\bb H}
\def\BC{\bb C}
\def\_\BC{\bbi C}
\def\RR{\bb R}

\def\ID{\mathbb I}
\def\SD{\rtimes}
\def\NN{\bb N}

\newcommand{\be}{\beta}
\newcommand{\ga}{\gamma}

\newcommand{\al}{\alpha}

\newcommand{\vth}{\vartheta}

\newcommand{\lga}{\longrightarrow}

\newcommand{\bra}{\begin{array}}
\newcommand{\era}{\end{array}}
\newcommand{\beq}{\begin{equation}}
\newcommand{\eeq}{\end{equation}}
\newcommand{\bqn}{\begin{eqnarray}}
\newcommand{\eqn}{\end{eqnarray}}
\newcommand\ben{\begin{enumerate}}
\newcommand\een{\end{enumerate}}
\newcommand\bei{\begin{itemize}}
\newcommand\ei{\end{itemize}}

\def\spt {space-time}
\def\gr {general relativity}

\def\minks {Minkowski spacetime}

\def\dS {de Sitter}
\def\ds {de Sitter}
\def\qft {quantum field theory}
\def\R{{\rm I\!R}}

\begin{document}

\maketitle

\abstract{We present a survey of rigourous quantization results obtained in 
recent works on quantum free fields in de Sitter space. For the 
``massive'' cases which are associated to principal series 
representations of the de Sitter group $SO_0(1,4)$, the construction is 
based on analyticity requirements on the Wightman two-point function. 
For the ``massless'' cases ({\it e.g.} minimally coupled or 
conformal), associated to the discrete series, the quantization 
schemes are of the Gupta-Bleuler-Krein type.}
\section{Introduction}

It is valuable to start out this review of recent results on de Sitter quantum field theory by quoting a sentence from the well-known Wald's monograph \cite{WALD} on QFT in curved space-time :
\vskip 0.5cm
\emph{It is worth noting that most of the available treatments of quantum field theory in curved spacetime 
either are oriented strongly toward mathematical issues (and deal, e.g., with $C^{\ast}$-algebras, KMS states, etc.) or are oriented toward 
a concrete physical problem (and deal, e.g., with particular mode function expansions of a quantum field in a certain spacetime). 
}
\vskip 0.5cm
De Sitter and Anti de Sitter \spt s play a fundamental role in 
cosmology, since they are, with \minks, the only maximally symmetric 
\spt ~solutions in \gr. Their 
respective invariance (in the relativity or kinematical sense) groups 
are the ten-parameters \ds ~$SO_{\scriptsize 0}(1,4)$ and anti \ds ~$SO_{\scriptsize 0}(2,3)$ groups.
Both may be seen as   deformations of the 
proper orthochronous Poincar\'e group $\RR^{1,3}\SD\, S0_{\scriptsize 
0}(1,3)$, the kinematical group of \minks.

The de Sitter [resp. anti-de Sitter]  \spt s are   solutions 
to the vacuum Einstein's equations with 
positive [resp. negative] cosmological constant $\Lambda$. This 
constant is linked to the (constant) Ricci  curvature $4 \Lambda$ of these 
\spt s. The  corresponding fundamental length is given by $H^{-1}:=\sqrt{3/(c\Lambda \def\dotx {\dot{x}})}$ 

Serious reasons back up any interest in studying Physics in 
such constant curvature spacetimes with maximal symmetry. The first 
one is the simplicity of their geometry, which makes consider them as 
an excellent laboratory model in view of studying Physics in more 
elaborate universes, more precisely with the purpose to set up a quantum field theory as much rigorous as possible \cite{ISH,[BD],FUL,WALD}. 

Higher dimensional Anti de Sitter spaces have becoming in the last 
years very popular because of their regularizing geometries. For 
instance they play an important role in some  versions of string or branes 
theories, and constitute presently the only cosmological example of 
the holographic conjecture.

Recent calculations \cite{lamizo}  suggested that the \dS ~solution may play an universal 
role as an ``osculating'' manifold for \spt.

Since the beginning of 
the eighties,  the de Sitter space has been 
playing a much popular role in inflationary cosmological scenarii  
\cite{[LI]}, where it is assumed that the cosmic dynamics was 
dominated by a term acting like a cosmoloical constant. 
More recently , observations on far high redshift 
supernovae \cite{[PER]}, on galaxy clusters \cite{[PH]}, and on 
cosmic microwave background radiation \cite{[SAP]} suggested an 
accelerating universe. Again, this can be explained only with such a 
term. 

On a fundamental level,   matter and energy are of quantum nature.
But the usual \qft ~is designed in \minks. Many theoretical and 
observational arguments   
plead in favour of setting up a rigorous \qft ~in \dS, and compare with our familiar minkowskian \qft. 
As a matter of fact, the symmetry properties of the dS solutions may allow the 
construction of such a theory. 

Also, the study of \dS ~\spt ~offers a specific interest because of   the regularization 
opportunity afforded by the curvature parameter as a ``natural'' 
cutoff for infrared or other divergences. 

On the other hand, as it will appear here, some of our most familiar concepts like time, energy, 
momentum, etc, disappear. They really need a new conceptual approach in 
de Sitterian relativity. However, it should be stressed that the current estimate on the cosmological constant does not allow any palpable experimental effect on the level of  high energy physics experiments, unless, as is explained in \cite{gano}, we deal with theories involving assumptions of infinitesimal masses like photon or graviton masses. We will tell more about this throughout the paper.

To summarize, 
the interest of setting up a QFT in de Sitter spacetime stems from 
\begin{itemize}
\item dS is  \textit{maximally symmetric}
\item Its symmetry is a one-parameter (curvature) deformation of minkowskian symmetry
\item It is so an excellent laboratory for both, mathematical or concrete, approaches to QFT
\item As soon as a constant curvature is present (like the currently observed one!), we lose some of our so familiar conservation laws like energy-momentum conservation.
\item What is then the physical meaning of a scattering experiment (``space" in dS is like the sphere $S^3$, let alone the fact that time is ambiguous)? 
\item Which relevant ``physical'' quantities are going to be considered as (asymptotically?, contractively?) experimentally
available?
\end{itemize}
The recent results on de Sitter Quantum Field Theory which we would 
like to report here can be viewed as a part of this program 
of understanding physics in the de Sitter universe. Of course, a huge 
amount of work has been done on de Sitter both on a classical and a 
quantum level since the Einstein's cosmological ``mistake'' and the 
first geometrical studies by de Sitter himself. For reasons that will 
become clear below, present  results are 
divided into two categories. The first category   concerns the 
``massive'' fields, so called for having Poincar\'e massive limits at 
null curvature. They have been developped in \cite{[BGM], [BM], 
[BEM], [Tak], [GT1], [BGMT], [GT2]}, and they are essentially 
characterised by analyticity properties of their Wightman two-point 
functions. The other category, developped in \cite{[DR], [GRT], 
[REN], [GT3], [GT4]}, deals with massless fields and other relevant 
fields, which require non standard quantization procedures. Both 
categories have a strong group theoretical flavor since they share, 
as a common obvious constraint, de Sitter covariance.

After describing the de Sitter geometry and kinematics (space and 
group) in Section 2, we  give in Section 3 the complete list of 
unitary irreducible representations of the de Sitter group and their 
possible contractive relations with the Wigner Poincar\'e 
representations. Then we  review in Sections 4 and 5 the main 
points of de Sitter QFT, pertaining to axiomatics as well as to 
technicality and problematic. Section 4 is devoted to de Sitter 
QFT for principal series (or ``massive'') fields based on the 
Wightman two-point function. We  explain in Section 5, through 
the example of the ``massless'' minimally coupled quantum field, how 
a new quantization, based on a Krein space of solutions of the de 
Sitter wave equation, allows to successfully deal with other fields 
like those pertaining to the discrete series. 

\section{De Sitter geometry and kinematics} 
                      
\subsection{The hyperboloid}

The de Sitter metric is the unique solution of the cosmological  vacuum 
Einstein's equation with positive cosmological constant $\Lambda=3~H^{2}$ (in units $c=1$). 
\bqn
\label{eins}
R_{\mu \nu} - \frac{1}{2} R g_{\mu \nu} + \Lambda g_{\mu \nu} = 0,\\ 
\nonumber R= R_{\mu \nu}g^{\mu \nu}= 4 \Lambda \equiv 12H^2. \eqn

\begin{figure}[h] 
   \centering
   \includegraphics[height=8cm,width=14cm]{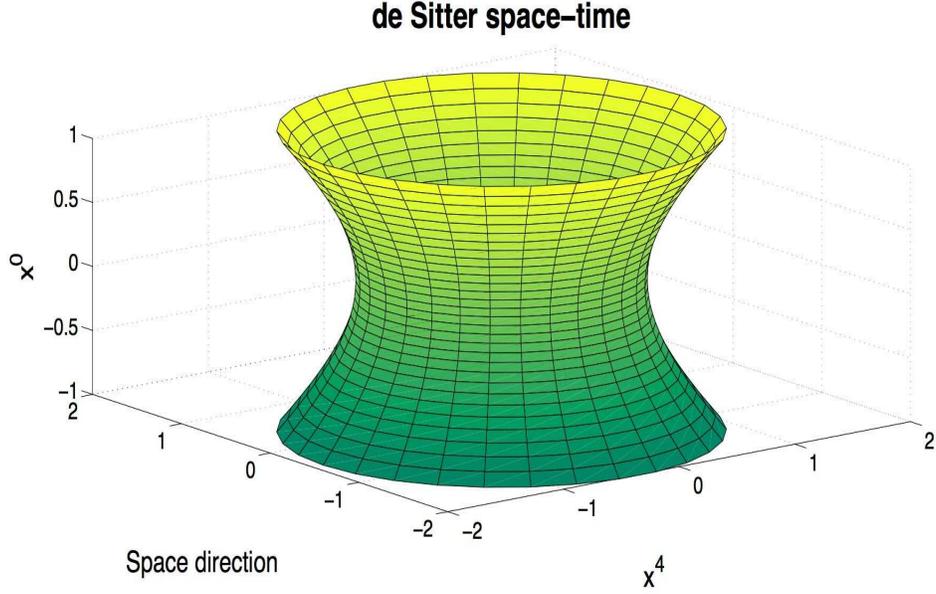} 
   \caption{de Sitter space-time as a hyperboloid  embedded in a five-dimensional Minkowski  space}
   \label{dSfig}
\end{figure}

The corresponding de Sitter space is conveniently seen as an 
one-sheeted hyperboloid (Fig \ref{dSfig}) embedded in a five-dimensional Minkowski 
space (the bulk):
\beq 
\label{dshyp}
M_H \equiv \{x \in \R^5 ;~x^2=\eta_{\alpha\beta}~ x^\alpha 
x^\beta =-H^{-2}\},\;\; 
\alpha,\beta=0,1,2,3,4, \eeq
where $\eta_{\alpha\beta}=$diag$(1,-1,-1,-1,-1)$. 

We can, for instance, adopt the following system of global coordinates :
 \begin{align}
 \label{coordh}
\nonumber x & := (x^0, \vec{x}, x^4)\\
 \nonumber x^{0}& =H^{-1}~\sinh(H\tau )\\
\nonumber \vec{x}&=(x^{1}, x^{2}, x^{3})=H^{-1}~\cosh(H\tau)~\sin(H\rho)~\vec{n}\\
x^{4}&=H^{-1}~\cosh(H\tau)~\cos(H\rho)
\end{align}
where $\vec{n}$ is a spatial 
direction, i.e., a spatial unit vector of $\R ^{3}$, and $R=H^{-1}$
for a point of the hyperboloid.

There is a global causal ordering on
the de Sitter manifold which is induced from that of the ambient 
spacetime ${\RR}^{5}$: given two events $x,y\in M_{H}$, one says that 
$x\geq y $ iff $ x-y\in\overline {V^+}$, where $\overline 
{V^{+}}=\{x\in{\RR}^{5}: x\cdot x \geq 0,\ {\rm sgn}\, x^{0}={+}\}$ 
is the future cone in ${\R}^{5}$. 

The closed causal future (resp. past) cone of a given point $x$ in 
$X$ is therefore the set $\{y\in M_H: y\geq x\}$ (resp. $\{y\in M_H: 
y\leq x\}$). Two events $x,y\in M_H$ are said in ``acausal relation" 
or ``spacelike separated" if they belong to the intersection of the 
complements of such sets, {\it i.e. } if $(x-y)^2 = -2(H^{-2} + 
x\cdot y) < 0$. 

\subsection{The de Sitter group}
           
The de Sitter relativity group is $G=SO_{0}(1,4)$, i.e. the component 
connected to the identity of the ten-dimensional pseudo-orthogonal 
group $SO(1,4)$. A familiar realization of the Lie algebra is that 
one generated by the ten Killing vectors
\beq
\label{kil}
K_{\alpha \beta} = x_{\alpha}\partial_{\beta} - 
x_{\beta}\partial_{\alpha}. \eeq
It is worthy to notice that there is no globally time-like Killing 
vector in de Sitter, the adjective time-like (resp. space-like) 
referring to the Lorentzian   four-dimensional 
 metric induced by that of the bulk. 

The universal covering of the de Sitter group is the symplectic 
$Sp(2,2)$ group, which  is needed when dealing with half-integer spins. 
It is suitably described as a subgroup of the group of $2\times2$ 
matrices with quaternionic coefficients: 
\beq
\label{sp} 
Sp(2,2) = \left\{ g= \left(
\bra{cc}
a & b \\
c & d \\
\era \right); \ a, b, c, d \in \HH , \ 
{\det}_{\scriptsize 4\times4}g = 1,
\ g^{\dagger}\gamma^0 g = \gamma^0 \right\}. \eeq
We recall that the group of quaternions $\HH \simeq \RR_{+}\times SU(2)$.  
We write   $(1\equiv e_4, ,e_i$ ($\simeq (-1)^{i+1}\sigma _{i})$ in $2\times 2$-matrix notations) the canonical basis for $\HH \simeq \RR^4$, with $i=1, 2,3$: any   quaternion will 
be written $q=(q^4, \vec{q})$ (resp. $q^{a}~e_a,~a=1,2,3,4$) in scalar-vector notations (resp. in euclidean metric notation). We also recall that the multiplication law is $qq' = (q^4 {q'}^4 - \vec{q}\cdot \vec{q'}, {q'}^4\vec{q} + {q}^4\vec{q'} + \vec{q}\times \vec{q'})$, the (quaternionic) conjugate of $q=(q^4, \vec{q})$ is $\bar{q}=(q^4, -\vec{q})$, the squared norm is $\Vert q \Vert^2 = q\bar{q}$, and the inverse of a nonzero quaternion is $q^{-1} = \bar{q}/ \Vert q \Vert^2$.

We have written  $g^{\dagger}= \bar{g}^{\scriptsize t}$ for  the quaternionic 
conjugate and transpose of the matrix $g$. The matrix
\beq
\label{gao}
\gamma^0= \left(
\bra{cc}
1 & 0 \\
0 & -1 \\
\era \right)
\eeq
is part of the Clifford algebra $\ga^{\al}\ga^{\be} + \ga^{\be} 
\ga^{\al} = 2 \eta^{\al \be}\ID$, the four other matrices having the 
following form in this quaternionic representation: 
\beq
\label{gaclif}
 \gamma^4= \left(
\bra{cc}
0 & 1 \\
-1 & 0 \\
\era \right) , \
\gamma^k= \left(
\bra{cc}
0 & e_k \\
e_k & 0 \\
\era \right), \ k= 1, 2, 3.\eeq
These matrices allow the following correspondence between points of 
the hyperboloid $M_H$ and $2\times 2$ quaternionic matrices of the 
form below:
\beq
\label{dsquat}
M_H \ni x \lga \barra{x} \equiv x^{\al}\ga_{\al} = \left( \bra{cc} 
x^0 & - {\cal P} \\
\overline{\cal P}& -x^0 \\
\era \right),
\eeq
where ${\cal P} \equiv (x^4, \vec{x}) \in \HH$. Note that we 
have
\begin{equation}
\label{screl}
x \cdot x= \barra{x} ^{\dag}\gamma ^{0}\barra{x}\gamma ^{0} ,~~x^{0}=
\frac{1}{4}\mbox{tr} \gamma^{0}\barra{x}.
\end{equation}The de Sitter 
action on $M_H$ is then simply given by
\beq
\label{dsact}
Sp(2,2) \ni g : \ \barra{x} \lga g\barra{x} g^{-1} = \barra{x}', \eeq
and this precisely realizes the isomorphism $SO_{\scriptsize 0}(1,4) 
\lga Sp(2,2)/\Z_2 $ through 
\beq
\label{dsact1}
SO_{\scriptsize 0}(1,4) \ni \Lambda(g) : \ x \lga \Lambda(g)x = x', \ 
\Lambda^{\al}_{\be} = \frac{1}{4} 
\mbox{tr}(\ga^{\al}g\ga_{\be}g^{-1}). \eeq

Another way to understand this group action on de Sitter is to resort 
to a specific (nonglobal) factorization of the group one can call 
space-time factorization and which is based on the group involution 
$g\lga \vartheta (g) = \ga^0\ga^4 g^{\dagger}\ga^0\ga^4$: 
\beq
\label{dsfact}
g = jl,\ j=\left(
\bra{cc}
\eta & 0 \\
0 & \bar{\eta} \\
\era \right)\left(
\bra{cc}
\cosh{\frac{\psi}{2}} & \sinh{\frac{\psi}{2}} \\ 
\sinh{\frac{\psi}{2}} & \cosh{\frac{\psi}{2}} \\ \era \right), \ 
l=\left(
\bra{cc}
\zeta & 0 \\
0 & \zeta \\
\era \right)\left(
\bra{cc}
\cosh{\frac{\varphi}{2}} & \hat{u}\sinh{\frac{\varphi}{2}} \\ 
-\hat{u}\sinh{\frac{\varphi}{2}} & \cosh{\frac{\varphi}{2}} \\ \era 
\right),
\eeq
where $ \psi, \varphi \in \RR, \ \zeta, \eta, \hat{u}= 
-\bar{\hat{u}}$ (``pure'' vector quaternion) $
\in SU(2)$. The factor $l$ is element of the (Lorentz) subgroup $L = 
\{l\in Sp(2,2);\ l\vartheta (l)=\ID\} \simeq SL(2, \BC)$ and the 
parameters $\zeta, \hat{u}, \varphi$ have the meaning of space 
rotation, boost velocity direction and rapidity respectively. The 
factor $j$ is a kind of ``space-time'' square root since we have
 \beq
 \label{dssqrt}
j \vartheta (j) = \left(
\bra{cc}
 \eta^2 \cosh{\psi} & \sinh{\psi} \\
 \sinh{\psi}& \overline{\eta^2} \cosh{\psi} \\ \era \right) 
\equiv\left( \bra{cc}
x^0 & -{\cal P} \\
\overline{\cal P}& - x^0 \\
\era \right)\left( \bra{cc}
0 & 1 \\
-1& 0 \\
\era \right) = \barra{x}\ga^4,
\eeq
where the equivalence holds modulo a determinant factor. We thus see 
that the group action (\ref{dsact}) is directly issued from the left action 
of the group on the coset $G/L$ through $j \lga gj= j'l'$. The 
Lorentz subgroup L is actually the stabilizer of $H^{-1}\ga^0\ga^4$. 
The latter corresponds to the point $O_H=(0,0,0,0,H^{-1})$ chosen as 
origin of the de Sitter universe, and $j$ maps this origin to the 
point $(x^0, {\cal P})$ in the notations (\ref{dsquat}). Note that the set 
$\{\psi, \eta^2\}$ in (\ref{dssqrt}) provides, through $x^0 =  \sinh{\psi}, {\cal P} =  \eta^2 \cosh{\psi}$, the system of global coordinates (\ref{coordh})
for $M_H$.

De Sitterian classical mechanics is understood along the traditional 
phase space approach. By phase space for an elementary system 
in de Sitter universe, we mean an orbit of the coadjoint 
representation of the group. We know that such an orbit is a 
symplectic manifold, and, as an homogeneous space, is homeomorphic to 
an even-dimensional group coset $Sp(2,2)/H_S$, where $H_S$ is the 
stabilizer subgroup of some orbit point. As a matter of fact, a 
scalar ``massive'' elementary system in de Sitter corresponds to the 
coset $Sp(2,2)/H_S$ where the subgroup
$H_S$ is made up with ``space'' rotations and ``time'' translations in 
agreement with the space-time factorization (\ref{dsfact}) of
$Sp(2,2)$:
\beq
 \label{isot}
H_S= \left\{ g= \left(
\bra{cc}
\zeta & 0 \\
0 & \zeta \\
\era \right) \left(
\bra{cc}
\cosh{\frac{\psi}{2}} & \sinh{\frac{\psi}{2}}\\ \sinh{\frac{\psi}{2}} 
& \cosh{\frac{\psi}{2}} \\ \era \right), \ \zeta \in SU(2), \ \psi 
\in \RR \right\}. \eeq

\section{De Sitter UIR and their physical interpretation }
Specific quantization procedures \cite{souriau,[KI],[GAPIE]} applied to the above classical phase 
spaces leads to their quantum counterparts, namely the quantum 
elementary systems associated in a biunivocal way to the the UIR's of 
the de Sitter group $Sp(2,2)$. Let us give a complete classification 
of the latter, following the works by Dixmier \cite{[DIX]} and 
Takahashi \cite{[TA]}. We recall that the ten Killing vectors (\ref{kil}) 
can be represented as (essentially) self-adjoint operators in Hilbert 
space of (spinor-)tensor valued functions on $M_H$, square integrable 
with respect to some invariant inner product, more precisely of the 
Klein-Gordon type. These operators take the form
  \beq
  \label{genrep}
K_{\al \be} \lga L_{\al \be} = M_{\al \be} + S_{\al \be}, 
  \eeq
where the orbital part is $M_{\al \be}=-i(x_{\al}\partial_{\be} - 
x_{\be}\partial_{\al})$ and the spinorial part $S_{\al \be}$ acts on 
the indices of functions in a certain permutational way. There are 
two Casimir operators, the eigenvalues of which determine completely 
the UIR's. They read: 
\beq
\label{cas1}
Q^{(1)} = - \frac{1}{2} L_{\al \be}L^{\al \be}, \eeq 
with eigenvalues $-p(p+1) - (q+1)(q-2)$ and
\begin{equation}
\label{cas2}
Q^{(2)} = - W_{\al} 
W^{\al}, \ W_{\al} = - \frac{1}{8}\epsilon_{\al \be \ga \delta \eta} 
L^{\be \ga}L^{\delta \eta},    \end{equation}with eigenvalues $-p(p+1)q(q-1)$.
Therefore, one must distinguish between
\bei
\item{\bf The discrete series} $\Pi^{\pm}_{p,q}$, \\ 
defined by $p$ and $q$ having integer or half-integer values, $p \ge q$. Note that 
$q$ may have a spin meaning.

Here, we must again distinguish between
\bei
\item {\it The scalar case} $\Pi_{p,0}$, $p=1,2, \cdots$; hereafter 
we refer to it as {\bf Dsc};  
\item {\it 
The spinorial case} $\Pi^{\pm}_{p,q}$, $q>0$, $p= \frac{1}{2}, 1, 
\frac{3}{2}, 2, \cdots$, $q=p, p-1, \cdots, 1$ or $\frac{1}{2}$:{\bf 
Dsp} \ei

\item {\bf The principal and complementary series} $\NU_{p,\sigma}$, 
\\ where $p$ has a spin meaning.
We put $\sigma =q~(1-q)$ which gives $q = \frac{1}{2}\left( 1 \pm \sqrt{1 - 4 \sigma^2} \right)$.

Like in the above, one distinguishes 
between
\bei
\item {\it The scalar case} $\NU_{0,\sigma}$, where \bei
\item $-2<\sigma < \frac{1}{4}$ for the complementary series: {\bf 
Cscm}, {\bf Csc0} for $\sigma=0$;
\item  $\frac{1}{4} \leq \sigma$ for the principal series: {\bf Pscm}. \ei
\item {\it The spinorial case} $\NU_{p,\sigma}$, $p>0$, where \bei
\item $0<\sigma < \frac{1}{4}$, $p=1,2, \cdots$, for the 
complementary series: {\bf Cspm}, \item $\frac{1}{4} \leq \sigma$, $p=1,2, 
\cdots$, for the integer spin principal series: {\bf Pspm}, \item $\frac{1}{4} < 
\sigma$, $p= \frac{1}{2}, \frac{3}{2}, \frac{5}{2} \cdots$, 
for the half-integer spin principal series: {\bf Pspm}. \ei
\ei
\ei

{\subsection{Contraction limits}

An important question to be addressed concerns the interpretation of these 
UIR's (or quantum de Sitter elementary systems) from a Minkowskian 
point of view. We mean by this the study of the contraction limit
$H\to 0$ of these representations, which is the quantum counterpart 
of the following 
geometrical and group contractions
\bei
\item $\lim _{ H \to 0} M_H = M_0$, the Minkowski spacetime tangent 
to $M_H$ at, say, the de Sitter origin point $O_H$,
\item $\lim _{ H \to 0}Sp(2,2) = {\cal P}^{\uparrow}_{+} (1, 3) = M_0 
\SD SL(2,\BC)$, the Poincar\'e group.
\ei
As a matter of fact, the ten de Sitter Killing vectors (\ref{kil}) contract 
to their Poincar\'e counterparts $K_{\mu \nu}$, $\Pi_{\mu}$, $\mu = 
0, 1, 2, 3$, after rescaling the four $K_{4\mu} \lga \Pi_{\mu} = H 
K_{4\mu} $. 

Now, we have to distinguish between the Poincar\'e massive and 
massless cases. We shall denote by ${\cal P}^{\stackrel{>}{<}}(m,s)$ 
the positive (resp. negative) energy Wigner UIR's of the Poincar\'e 
group with mass $m$ and spin $s$. For  interesting discussion and precision on this confusing notion of mass in ``desitterian Physics'', we will give below details on the  work by Garidi \cite{GAR1}. We shall make use of similar 
notation
${\cal P}^{\stackrel{>}{<}}(0,s)$ for the Poincar\'e massless case 
where $s$ reads for helicity. In the latter case, conformal 
invariance leads us to deal also with the discrete series 
representations
(and their lower limits) of the (universal covering of the) 
conformal group or its double covering $SO_0(2,4)$ or its fourth 
covering $SU(2,2)$. These UIR's are denoted in the sequel by
${\cal C}^{\stackrel{>}{<}}(E_0,j_1, j_2)$, where $(j_1,j_2) \in 
\NN/2 \times \NN/2$ labels the UIR's of $SU(2) \times SU(2)$ and 
$E_0$ stems for the positive (resp. negative) conformal energy. The 
de Sitter contraction limits can be summarized in the following 
diagrams.

\paragraph{Massive case} Solely the principal series representations  {\bf 
Pscm} and {\bf Pspm} are 
involved here (from where comes the name of de Sitter ``massive 
representations''). Introducing the parameter $\nu $ through $\sigma 
= \nu^2 + 1/4$, and the Poincar\'e mass $m=\nu H$, we have 
\cite{[MINI],GAHURE} 
\beq
\label{contr}
{\NU}_{s,\sigma} \lga_{H\to 0, \nu \to \infty} {c_>\cal P}^{>}(m,s) 
\oplus c_<{\cal P}^{<}(m,s),
\eeq
where one of the ``coefficients'' among $c_<, c_>$ can be fixed to 1 whilst the other one will vanishes.
Note here the evidence of the energy ambiguity in de Sitter relativity, 
exemplified  by the possible breaking of dS irreducibility into a direct sum of 
two Poincar\'e UIR's with positive and negative energy respectively. 
This phenomenon is linked to the existence in the de Sitter group of 
a specific discrete symmetry, precisely $\ga_0 \in Sp(2,2)$, which 
sends any point $(x^0, {\cal P}) \in M_H$ (with the notations of 
(2.7)) into its mirror image $(x^0, -{\cal P}) \in M_H$ with respect 
to the $x^0$-axis. Under such a symmetry the four generators 
$L_{a0}$, $a = 1,2,3,4$, (and particularly $L_{40}$ which contracts 
to energy operator!) transform into their respective opposite 
$-L_{a0}$, whereas the six $L_{a b}$'s remain unchanged. 

Note that the well-known ambiguity concerning the existence of a vacuum (``$\alpha$-vacua'') in de Sitter quantum field theory  originates in the above contraction arbitrariness. 

In the context of the notion of mass in ``desitterian Physics'', 
the following  ``mass'' formula has been proposed by Garidi \cite{GAR1} in terms of the dS RUI parameters $p$ and $q$:
\begin{equation}
\label{garidimass}
m^2_H= \langle Q^{(1)} \rangle_{\mathrm{dS}} - \langle Q^{(1)}_{p=q} \rangle_{\mathrm{dS}}= [(p - q)(p + q - 1)] \hbar^2H^2/c^4.
\end{equation}
 This formula is natural in the sense that when the second-order wave equation
\begin{equation*}
\label{dswaveeq}
\left(Q^{(1)} - \langle Q^{(1)} \rangle_{\mathrm{dS}}\right)\varphi = 0,
\end{equation*}
obeyed by rank $r$ tensor fields carrying a dS UIR, is written in terms of the Laplace-Beltrami operator
$\Box_H$ on the dS manifold, one gets (in units $\hbar = 1 = c$)
\begin{equation}
\label{dswaveeq}
\left(\Box_H  + H^2 r(r+2) + H^2 \langle Q^{(1)} \rangle_{\mathrm{dS}}\right)\varphi = 0.
\end{equation}
Moreover, the minimal value assumed by  the eigenvalues of the first Casimir in the set of RUI in the discrete series is precisely reached at $p=q$, which corresponds to the ``conformal'' massless case.
  The Garidi mass has the advantages to encompass all  mass formulas introduced within a de-sitterian context, often in a purely mimetic way in regard with their minkowskian counterparts.

Actually,  given a minkowskian mass $m$ and a ``universal'' length $R =:\sqrt{3/\vert \Lambda\vert}= c\,H^{-1}$ (we here restore for a moment all physical units) , nothing prevents us to consider the dS UIR parameter $\nu$ (principal series), specific of a ``physics'' in constant-curvature space-time, as a meromorphic functions of the dimensionless physical (in the minkowskian sense!) quantity, expressed in terms of  various other quantities introduced in this paper,
\begin{equation}
\label{nodimm}
 \vth \equiv  \vth_m \stackrel{\mathrm{def}}{=} \frac{\hbar}{Rmc}
= \frac{\hbar \sqrt{\vert \Lambda} \vert}{\sqrt{3}mc}  = \frac{\hbar\, H}{ mc^2}.
\end{equation}

We give in Table \ref{tab:table1}  the values assumed by the quantity $\vth$ when $m$ is taken as  some known masses and $\Lambda$ (or $H_0$) is given its present day estimated value. We  easily  understand from this table that the currently estimated value of the cosmological constant has no practical effect on our familiar massive fermion or boson fields. Contrariwise, adopting  the  de Sitter point of view appears as inescapable when we deal with  infinitely small masses, as is done in standard inflation scenario. 

\begin{table}[h]
  \centering 
 { \begin{tabular}[c]{|c|c|}\hline
Mass $m$ & $\vth_m \approx $\\ \hline
$m_{\Lambda}/\sqrt{3}\approx 0.293\times 10^{-68}$kg & 1 \\ \hline
up. lim. photon mass $m_{\gamma}$& $0.29 \times 10^{-16}$\\ \hline
up. lim. neutrino mass $m_{\nu}$& $0.165 \times 10^{-32}$\\ \hline
electron mass $m_e$& $0.3 \times 10^{-37}$\\ \hline
proton mass $m_p$& $0.17  \times 10^{-41}$ \\ \hline
$W^{\pm}$  boson mass & $0.2  \times 10^{-43}$\\ \hline
Planck mass $M_{Pl}$& $0.135  \times 10^{-60}$\\ \hline
\end{tabular}}
\label{tab:table1}
\caption{Estimated values of the dimensionless physical quantity $\vth \equiv \vth_m =:  \frac{\hbar \sqrt{\vert \Lambda} \vert}{\sqrt{3}mc}=  \frac{\hbar\, H}{ mc^2} \approx 0.293\times 10^{-68}\times m_{\mathrm{kg}}^{-1} $ for some known masses $m$ and the present day estimated value of the Hubble length $ c/H_0 \approx 1.2\times 10^{26} \mathrm{m}$ \cite{pada}.}
\end{table}

Now, we may consider the following Laurent expansions of the principal series dS UIR parameter $\nu$  in a certain neighborhood of $\vth = 0$:
\begin{equation}
\label{laurentnu}
 \nu =     \nu (\vth)=   \frac{1}{\vth} + e_0 + e_1 \vth + \cdots e_n \vth^n + \cdots,  \  \vth \in (0, \vth_1) \ \mbox{convergence interval},
\end{equation}
Coefficients $e_n$ are pure numbers to be determined.
We should be aware that nothing is changed in the  contraction formulas from the point  of view of a minkowskian observer, except that we allow to consider  positive as well as negative values of $\nu$ in a (positive) neighborhood of $\vth = 0$: multiply this formula by $\vth$ and go to the limit $\vth \to 0$. 
We recover asymptotically the relation
\begin{equation}
\label{dsnumass1}
m= \vert \nu \vert  \hbar H/c^2 = \vert  \nu \vert  \frac{\hbar}{c} \sqrt{\frac{ \vert \Lambda\vert}{3}}.
\end{equation}
As a matter of fact, the Garidi mass is perfect example of such an expansion since it can be rewritten as the following expansion in the parameter $\vth \in ( 0,1/\vert s- 1/2 \vert]$:
\begin{align}
\label{garmassnu}
  \nonumber \nu =  \sqrt{\frac{1}{\vth^2} - (s-1/2)^2}\\
  &= \frac{1}{\vth} -(s-1/2)^2\left(\frac{\vth}{2} + O(\vth^2)\right),  
\end{align}
Note the particular symmetric place occupied by the spin $1/2$ case with regard to the scalar case $s=0$ and the boson case $s=1$. More details concerning this discussion are given in \cite{gano}.

\paragraph{Massless (conformal) case} Here we must distinguish between the 
scalar massless case, which involves the unique complementary series 
UIR $\NU_{0,0}$ to be contractively Poincar\'e significant, and the 
spinorial case where are involved all representations 
$\Pi^{\pm}_{s,s}, \ s>0$ lying at the lower limit of the discrete 
series. The arrows $\hookrightarrow $ below designate unique 
extension.
\bei
\item Scalar massless case : {\bf Csc0}.
\beq \left. \begin{array}{ccccccc}
&	& {\cal C}^{>}(1,0,0)
& &{\cal C}^{>}(1,0,0) &\hookleftarrow &{\cal P}^{>}(0,0)\\ 
\NU_{0,0} &\hookrightarrow & \oplus
&\stackrel{H=0}{\longrightarrow} & \oplus & &\oplus \\ 
&	& {\cal C}^{<}(-1,0,0)&
& {\cal C}^{<}(-1,0,0) &\hookleftarrow &{\cal P}^{<}(0,0),\\ 
\end{array} \right. \eeq

\item Spinorial massless case : {\bf Dsp0}.
\beq \left. \begin{array}{ccccccc}
&	& {\cal C}^{>}(s+1,s,0)
& &{\cal C}^{>}(s+1,s,0) &\hookleftarrow &{\cal P}^{>}(0,s)\\ 
\Pi^+_{s,s} &\hookrightarrow & \oplus
&\stackrel{H=0}{\longrightarrow} & \oplus & &\oplus \\ 
&	& {\cal C}^{<}(-s-1,s,0)&
& {\cal C}^{<}(-s-1,s,0) &\hookleftarrow &{\cal P}^{<}(0,s),\\ 
\end{array} \right. \eeq
\beq \left. \begin{array}{ccccccc}
&	& {\cal C}(s+1,0,s)
& &{\cal C}^{>}(s+1,0,s) &\hookleftarrow &{\cal P}^{>}(0,-s)\\ 
\Pi^-_{s,s} &\hookrightarrow & \oplus
&\stackrel{H=0}{\longrightarrow} & \oplus & &\oplus \\ 
&	& {\cal C}^{<}(-s-1,0,s)&
& {\cal C}^{<}(-s-1,0,s) &\hookleftarrow &{\cal P}^{<}(0,-s),\\ 
\end{array} \right. \eeq

\ei

Finally, all
other representations have either non-physical Poincar\'e contraction 
limit or have no contraction limit at all.

\section{Quantum field theory in de Sitter space: the ``massive'' 
case} Let us first outline the main features of a quantum field 
theory on de Sitter based on the properties of the Wightman 
functions. For free fields whose the one-particle sector is 
determined by a given de Sitter UIR in the principal and the 
complementary series, one resorts to an axiomatic {\it \`a la } 
Wightman \cite{[SW]}, where precisely the so-called two-point 
Wightman function is required to satisfy the following four criteria. 
\ben
\item[(i)] Covariance with respect to the given UIR. \item[(ii)] 
Locality/(anti-)commutativity, wich respect to the causal de Sitter 
structure. \item[(iii)] Positive definiteness (Hilbertian Fock 
structure). \item[(iv)] Normal (maximal?) analyticity. \een
Then the field itself can be reobtained from the Wightman function 
via a Gelfand-Na\"{\i}mark-Segal (G.N.S.) type construction. Note that (i),(ii), and (iii) are 
analogous to the Minkowskian QFT requirements. On the other hand, 
Condition (iv) will play the role of a spectral condition in the 
absence of a global energy-momentum interpretation in de Sitter. This 
condition implies a thermal Kubo-Martin-Schwinger (K.M.S.) interpretation.

For ``generalized'' free fields, the theory is still encoded entirely 
by a two-point function: all truncated $n$-point functions, $n>2$, 
vanish, as does the ``1-point" function. The axiomatic imposes the 
2-point functions to obey the same conditions (i)-(iv), apart from 
the fact that a certain not necessarily irreducible unitary 
representation is now involved. However, the Plancherel content of 
this involved UR should be restricted to the principal series, 
and this decomposition allows a 
K\"{a}llen-Lehman type representation of the 2-point function. 
Finally, for interacting fields in dS, the \underline{set} of 
$n$-point functions is assumed to satisfy
\ben
\item[(i)] Covariance with respect to a certain dS unitary representation. \item[(ii)] 
Locality/(anti-)commutativity. \item[(iii)] Positive definiteness.
\item[(iv)]``Weak'' spectral condition in connection with some 
analyticity requirements. \een

\subsection{Plane waves} 

We refer to \cite{[BGM], [BM], [BEM]} for details. 

We consider the eigenvector  equations of the second-order Casimir 
operator  for the principal and complementary series. 
For any eigenvalue, they give a  Klein-Gordon-like or Dirac-like 
  equation. The whole quantum 
field construction rests upon those elementary pieces which are the 
so-called dS plane wave solutions. Let us 
here recall those equations : \bei
\item Principal series ({\bf Pscm} and {\bf Pspm}): $\NU_{p=s, \sigma = \nu^2 + \frac{1}{4}}$: 
\begin{equation}
[Q^{(1)}_s - (\nu^2 + \frac{9}{4} - s(s+1))]~\psi(x) = 0,\end{equation} where 
$\nu \geq 0$ for $s=0, 1, 2,
\cdots$, and $\nu > 0$ for $s= \frac{1}{2}, \frac{3}{2},\cdots$. 
\item Complementary series ({\bf Cscm} and {\bf Cspm}): $\NU_{p=s,\sigma}$: \begin{equation}
[Q^{(1)}_s - (\sigma + 2 - s(s+1))]~\psi(x) = 0,\end{equation} where $ -2 < \sigma < \frac{1}{4}$ for 
$s=0$, and $0<\sigma <\frac{1}{4}$ for $s= 1,2,\cdots$. \ei
The de Sitter plane waves have the general form \beq
\psi(x) = {\cal D}(\xi, z) (z\cdot \xi)^{\mu} \mid_{z=x}, \eeq
where
\ben
\item[$\rightarrow$] ${\cal D}(\xi, z)$ is a vector-valued 
differential operator such that $\psi(x)$ is a relevant tensor-spinor 
solution of the wave equation. 
\item[$\rightarrow$] The vector $\xi = 
(\xi^0,\vec{\xi},\xi^4) $ belongs to  $C^{\pm} = \{ \xi \in \RR^5: \ \xi\cdot \xi = 
0, \ \mbox{sgn} (\xi^0) = \pm \}$, the ``future'' null cone in the 
ambient space $\RR^5$. This vector  $\xi$ plays the 
role of a four-momentum. Note that $(z \cdot \xi)^{\mu}$ is a 
$\Box _{5}$-harmonic function in $1+4$ Minkowski.
\item[$\rightarrow$] The complex five-vector 
$z$ belongs to the tubular domains ${\cal T}^{\pm}$: ${\cal T}^{\pm} 
= (\RR^5 \pm i V^{+}) \cap M^{(c)}_H$, where $M^{(c)}_H$ is the 
complexification of the dS hyperboloid $M_H$ and $\RR^5 \pm i V^{+}$ 
are the forward and backward tubes in $\BC^5$.
\item[$\rightarrow$] The complex power $\mu$ is such that $\psi$ is 
solution to the wave equation. \een
The occurrence of complex variables in these expressions is not 
fortuitous. It is actually at the heart of the analyticity 
requirements (iv), as will appear through the following explicit 
examples. 

\subsection{The example of the scalar case $s=0$ ({\bf Pscm} and {\bf Pspm})} 
                
The complex plane waves are given by 
\begin{equation}
\label{wapla}
\psi(z) = (Hz\cdot \xi)^{\mu},~\mbox{where}~
\mu = -\frac{3}{2} + i \nu, \ \nu 
\in \RR\ ,         
\end{equation}
for the principal series (p.s.) 
$\NU_{0, \sigma = \nu^2 + \frac{1}{4}}$ and
\begin{equation}
-3< \mu = -\frac{3}{2} \pm \sqrt{1-2\sigma} < 0 \     
\end{equation}
for the 
complementary series (c.s.) $\NU_{0, \sigma }$.

The term {\it wave plane} in the case of the principal series is 
consistent with the null curvature limit (\ref{contr}).

We use the  parametrization (\ref{coordh})
  of the hyperboloid. At the $R \rightarrow \infty$ limit, $x 
  \rightarrow(\tau, 
\rho~\vec{n},\infty)$, that we consider as the point   $X:=(X^{0}=\tau, 
\vec{X}=\rho ~\vec{n}) \in M^{1,3}$. To take the limit for the 
plane wave, we   write $m=H\nu$, leading to 
\beq
\label{limwpl}
\lim_{H\to 0} (Hx(X)\cdot \xi)^{-\frac{3}{2} +imH^{-1}} = 
\exp{ik\cdot X}, \eeq
where, in Minkowskian-like coordinates, $\xi = 
(\frac{k^0}{m},\frac{\vec{k}}{m},-1) \in C^{+}$, and\\
$x(X) = (H^{-1} \sinh{HX^0},
\vec{x} = H^{-1}\frac{\vec{X}}{\Vert \vec{X}\Vert }\cosh{HX^0} 
\sin{H\Vert \vec{X}\Vert}, x^4 = H^{-1}\cosh{HX^0} \cos{H\Vert 
\vec{X}\Vert})$. 

The two-point function is analytic in the tuboid ${\cal T}^{-} \times 
{\cal T}^{+}$ and reads (for the principal series)
\begin{align}
\label{comwight}
\nonumber     W_{\nu} (z_1,z_2) 
   = &  c_{\nu} \int_{\ga} (z_1\cdot \xi)^{-\frac{3}{2} 
+i\nu}(\xi\cdot z_2)^{-\frac{3}{2} -i\nu}\,d\mu_{\ga} (\xi)  \\
   = &\frac{H^2 \Gamma(\frac{3}{2} +i\nu) \Gamma(\frac{3}{2} -i\nu)}{2^4 
\pi^2} P^{5}_{-\frac{3}{2} +i\nu} (H^2 z_1\cdot z_2).
\end{align}
The integration is performed on an ``orbital basis'' $\ga \subset 
C^+$. The symbol $P^{\lambda}_{\al}$ stems for a generalized Legendre 
function, and the coefficient factor is fixed by the \emph{Hadamard condition}. We recall that the Hadamard condition imposes that the short-distance behavior of the two-point function 
of the field should be the same for Klein-Gordon fields on curved space-time as for corresponding Minkowskian free field.
  In case of dS (and many other curved space-times) it selects a unique vacuum state
In case of dS, this selected vacuum coincides with the \emph{euclidean} or \emph{Bunch-Davies} vacuum state
structure. 

The corresponding Wightman function ${\cal W}_{\nu}(x_1,x_2) = 
\langle \Omega,
\phi(x_1) \phi(x_2) \Omega \rangle$, where $\Omega$ is the Fock 
vacuum and $\phi$ is the field operator seen as an operator-valued 
distribution on $M_H$, is the boundary value {\it bv}$_{{\cal 
T}^{\mp} \ni z_{\stackrel{1}{2}}\to x_{\stackrel{1}{2}}} 
W_{\nu}(z_1,z_2)$. Its integral representation is given by:
\begin{align}
\label{realwight}
 \nonumber {\cal W}_{\nu}(x_1,x_2) =  & c_{\nu}\int_{\ga}((x_1\cdot \xi)_{+}^{-\frac{3}{2} +i\nu} + e^{-i\pi 
(-\frac{3}{2} +i\nu)}(x_1\cdot
\xi)_{-}^{-\frac{3}{2} +i\nu})((\xi\cdot x_2)_{+}^{-\frac{3}{2} 
-i\nu}   \\
  +  &  e^{-i\pi (-\frac{3}{2} -i\nu)}(\xi\cdot 
x_2)_{-}^{-\frac{3}{2} -i\nu})\,d\mu_{\ga} (\xi). 
\end{align}
This function satisfies all QFT requirements: \ben
\item[(i)] Covariance: ${\cal W}_{\nu}(\Lambda^{-1} x_1, \Lambda^{-1} 
x_2) = {\cal W}_{\nu}(x_1,x_2)$, for all $\Lambda \in SO_0(1,4)$. 
\item[(ii)] Local commutativity: ${\cal W}_{\nu}(x_1,x_2) = {\cal 
W}_{\nu}(x_2,x_1)$, for every space-like separated pair $(x_1,x_2)$.
\item[(iii)] Positive definiteness: $0 \leq \int_{M_H \times M_H} 
\overline{f}(x_1) {\cal W}_{\nu}(x_1,x_2) f(x_2) \, d\mu(x_1) \, 
d\mu(x_2)$ for any test function $f$, and where $ d\mu(x)$ is the 
$O(1,4)$ invariant measure on $M_H$. \item[(iv)] Maximal analyticity: 
$ W_{\nu}(z_1,z_2) $ can be analytically continued in the cut-domain 
$\Delta = (M_H^{(c)} \times M_H^{(c)})\setminus \Sigma^{(c)}$ where 
the cut is defined by $\Sigma^{(c)} = \{ (z_1,z_2) \in M_H^{(c)} 
\times M_H^{(c)}; \, (z_1-z_2)^2 =\rho,\, \rho \geq 0 \}$.
\een

\subsection{The example of the spinorial case $s=\frac{1}{2}$} 

The involved UIR is here $\NU_{\frac{1}{2}, \sigma = \nu^2 + 
\frac{1}{4}}$ \cite{[BGMT]}. For simplicity we shall put $H=1$ in the 
sequel. We now have four independent plane wave solutions: 
\beq 
\label{spwapl}
\psi^{(1)}_{r, \nu} = (Hz\cdot \xi)^{-2 +i\nu} u^{(1)}_r (\xi), \ 
\psi^{(2)}_{r, \nu} = (Hz\cdot \xi)^{-2 -i\nu} \barra{z} u^{(2)}_r 
(\xi), \ r=1,2, \ \xi \in C^{+}, \eeq
where the four 4-spinors $ u^{(1)}_r$, $u^{(2)}_r$ are independent 
solutions to $\barra{\xi} u(\xi) = 0$. The resulting $4\times 4$ 
two-point function is analytic in the tuboid ${\cal T}^-\times {\cal 
T}^+$ and is given by:
\bqn
\label{comsp}
\nonumber S^{\nu} (z_1,z_2) & = & a_{\nu} \int_{\ga} (z_1\cdot 
\xi)^{-2 +i\nu}(\xi\cdot z_2)^{-2 -i\nu} (\frac{1}{2} \barra{\xi} 
\ga^4)\,d\mu_{\ga} (\xi) \\ &= &\frac{1}{8} A_{\nu} \left\lbrack
(2-i\nu) P^{7}_{-2 - i\nu} ( z_1\cdot z_2) \barra{z}_1 -(2+i\nu) 
P^{7}_{-2 + i\nu} ( z_1\cdot z_2) \barra{z}_2 \right \rbrack,
\eqn
where $A_{\nu} = (i\nu (1+\nu^2))/8\pi \sinh{\pi \nu}$ is 
imposed by the Hadamard condition. The Wightman function 
$S^{\nu}(x_1,x_2) = \langle \Omega, \Psi(x_1) \otimes 
\overline{\Psi}(x_2) \Omega \rangle$, where the spinor field
$\Psi = (\Psi_i)$, $i=1,2,3,4$, and its adjoint $ \overline{\Psi} 
\equiv \Psi^{\dagger} \ga^0 \ga^4$ are operator-valued distributions 
on $M_H$, is the boundary value {\it bv}$_{{\cal T}^{\mp} \ni 
z_{\stackrel{1}{2}}\to x_{\stackrel{1}{2}}} S^{\nu}(z_1,z_2)$.
This function meets all axiomatic requirements: \ben
\item[(i)] Covariance: $gS^{\nu}(\Lambda^{-1}(g) x_1, \Lambda^{-1}(g) 
x_2)i(g^{-1}) = S^{\nu}(x_1,x_2)$, for all $g \in Sp(2,2)$. The group 
involution $i(g)$ is defined by $i(g) =-\ga^4 g \ga^4$.
\item[(ii)] Local \underline{anti}commutativity: $ 
S_{i\bar{j}}(x_1,x_2) = S'_{i\bar{j}}(x_1,x_2) \equiv - \langle 
\Omega, \overline{\Psi}_{\bar{j}}(x_2) \Psi_{i}(x_1) \Omega \rangle$, 
for every space-like separated pair $(x_1,x_2)$. \item[(iii)] 
Positive definiteness: $0 \leq \int_{M_H \times M_H} 
\overline{h}(x_1) S^{\nu}(x_1,x_2) h(x_2) \, d\mu(x_1) \, d\mu(x_2)$ 
for every 4-spinor valued test function $h$. \item[(iv)] Maximal 
analyticity: $ S^{\nu}(z_1,z_2) $ can be analytically continued in 
the cut-domain $\Delta = (M_H^{(c)} \times M_H^{(c)})\setminus 
\Sigma^{(c)}$. \een
Higher-spin QF cases, for the principal or the complementary series, 
are similar to the ones presented in the above, and we refer to 
\cite{[GT1], [GT2]} for details.

\section{``Massless'' minimally coupled quantum field} 

The so-called ``massless'' minimally coupled quantum field (which is 
\underline{not} ``massless'' in our sense, even though the corresponding Garidi mass exceptionally vanishes!) occupies under many 
aspects a central position in de Sitter theories (see \cite{[GRT], 
[REN]} and references therein). On the mathematical side, it is 
associated to the lowest limit, namely
$\Pi_{1,0}$, of the discrete series, and we shall see below some 
interesting features of this representation, like its place within a 
remarkable indecomposable representation. On the physical side, it 
has been playing a crucial role in inflation theories \cite{[LI]}, it 
is part of the Gupta-Bleuler structure (again an indecomposable UR is 
involved here!) for the massless spin 1 field (de Sitter QED, 
\cite{[GT3]}), and it is the elementary brick for the construction of 
massless spin 2 fields (de Sitter linear gravity
\cite{[GT4]}).

The wave equation for $\Pi_{1,0}$ is
\beq
\label{mmceq}
Q^{(1)} \psi (x) = 0 \ \Leftrightarrow \Box \psi (x) = 0, \eeq
where $\Box $ is the dS Laplace-Beltrami operator. ``Mode'' solutions 
$\phi_{Llm}$ to (\ref{mmceq}) are expressed in terms of the following bounded 
global coordinates (suitable for the compactified dS $\simeq$ Lie 
sphere $S^3\times S^1$):
\beq
\label{concoor}
x = (x^0 = H^{-1} \tan{\rho}, (\vec{x}, x^4) = \frac{u}{H\cos{\rho}}) 
\equiv (\rho, u), \ -\frac{\pi}{2} < \rho <\frac{\pi}{2}, \ u \in 
S^3. \eeq
The coordinate $\rho$ is timelike and plays the role of a conformal 
time, whereas $u$ coordinatizes the compact spacelike manifold. The 
``strictly positive'' modes are given by 
\beq
\label{mmcmod}
\phi_{Llm} (x) = A_L (L e^{-i(L+2)\rho} + (L+2)e^{-iL\rho}) {\cal 
Y}_{Llm} (u), \ L= 1, 2, \cdots, \ 0\leq l\leq L, \ 0\leq \vert m 
\vert \leq l, \eeq
where the ${\cal Y}_{Llm}$ are the spherical harmonics on $S^3$. 
These modes form an orthonormal system with respect to the 
Klein-Gordon inner product, 
\beq
\label{KGds}
\langle \phi, \psi \rangle = \frac{i}{\pi^2} \int_{\rho = 0} 
\bar{\phi}(\rho, u) \stackrel{\leftrightarrow}{\partial}_{\rho} 
\psi(\rho,u)\, du. \eeq
The normalisation constant $A_L = \frac{H}{2} \lbrack 2(L+2)(L+1)L 
\rbrack^{-1/2}$ breaks down at $L=0$: this is called the 
``zero-mode'' problem, and this problem is related to the fact that 
the space generated by the strictly positive modes (\ref{mmcmod}) is 
\underline{not} dS invariant. It is only $O(4)$ invariant. If one 
wishes to restore full dS invariance, it is necessary to deal with 
the $L=0$ solutions to (\ref{mmceq}). There are two of them, namely the 
constant ``gauge'' solution $\psi_g$ and the ``scalar'' solution 
$\psi_s$: 
\beq
\label{gasc}
\psi_g = \frac{H}{2\pi}, \ \psi_s = -\frac{iH}{2\pi}(\rho + 
\frac{1}{2} \sin{2\rho}). \eeq
Both are null norm states, and the constants are chosen in order to 
have $\langle \psi_g, \psi_s \rangle = 1$. Then we define the ``true 
'' normalized zero mode: 
\beq
\label{omod}
\phi_{000} = \psi_g + \frac{1}{2}\psi_s \equiv \phi_0 , \ \langle 
\phi_0, \phi_0 \rangle = 1. \eeq
Now, applying de Sitter group actions on it produces 
\underline{negative} ($\overline{\phi}_{Llm}$) as well as positive 
modes ($\phi_{Llm}$). We thus see an indefinite inner product space 
${\cal H}$ emerges under the form of a direct sum 
Hilbert$\oplus$anti-Hilbert. This is called a Krein space \cite{MINT,BOG}. More 
precisely, one defines the Hilbert space ${\cal H}_+$ generated by 
the positive modes (including the zero mode): 
\beq
\label{poshilb}
{\cal H}_+ = \{\sum_{(Llm) \equiv k \geq 0} c_k \phi_k;\ \sum_{k\geq 
0} \vert c_k \vert^2 < \infty \}.
\eeq
Similarly, one defines the anti-Hilbert space ${\cal H}_-$ as that 
one generated by the ``negative'' modes $\overline{\phi}_k$, $k\geq 
0$, or equivalently the conjugates of the positive ones. Note that 
$\langle \phi_k, \phi_{k'} \rangle = \delta_{kk'} = -\langle 
\overline{\phi}_k, \overline{\phi}_{k'} \rangle$. Then ${\cal H} = 
{\cal H}_+ \oplus {\cal H}_-$. This Krein space is de Sitter 
invariant, but its direct sum decomposition is \underline{not}. It 
has a Gupta-Bleuler triplet structure \cite{[GA]} which carries an 
indecomposable representation of the de Sitter group. The involved 
Gupta-Bleuler triplet is the chain of spaces \beq
\label{GBtri}
\BC \psi_g \equiv {\cal N} \subset \{ c_g\psi_g + \sum_{k> 0} c_k 
\phi_k\} \equiv {\cal K} \subset {\cal H}.
\eeq
Space ${\cal N}$ is a null norm space whereas ${\cal K}$ is a 
degenerate inner product space. The coset space ${\cal K}/{\cal N}$ 
\underline{is} the space of physical states, and it is precisely this 
Hilbert space which carries the UIR $\Pi_{1,0}$. {\it A contrario}, 
the coset space ${\cal H}/{\cal K}$ is the space of unphysical 
states. It is however an (anti) Hilbert space which carries also 
$\Pi_{1,0}$. Noticeing that the coset by itself of the space of 
constant functions or gauge states ${\cal N}$ carries the trivial 
representation $\NU_0$ (on which both Casimir operators vanish), the 
whole indecomposable representation carried by the Krein space can be 
pictured by \cite{[GA]}
\beq
\Pi_{1,0} \lga \Pi_{1,0} \lga {\NU}_0.
\eeq
Also note that this indecomposable structure is based on the exact 
sequence of carrier spaces \cite{[PI]}
\beq
\label{indec}
 \bra{ccccccc}
0 &\lga &{\cal N}& \stackrel{i}{\lga} &{\cal K}	&\lga &{\cal H} \\
&	&	&	& \downarrow	&	&\downarrow \\
&	&	&	&{\cal K}/{\cal N}&	&{\cal H}/{\cal K}\\
&	&	&	& \downarrow	&	&\downarrow \\
&	&	&	& 0	&	& 0
\era
\eeq
Let us turn to the quantization of this field. If we adopt the usual 
representation of the canonical commutation rules, namely if the 
quantized field $\varphi$ is given by 
 \beq
 \label{usccr}
\varphi = \sum_{k\geq 0}(A_k \phi_k(x) + h.c.), \ \ \ \ \lbrack A_k, 
A^{\dagger}_{k'} \rbrack = 2 \delta_{kk'},
\eeq
we get a QFT which is \underline{not} dS covariant: it is 
$SO(4)$-covariant only, and the so defined vacuum is solely 
$SO(4)$-invariant. In order to restore the full dS-covariance, one 
has to resort to the following new representation of the {\bf ccr} 
\beq
 \label{newccr}
\varphi = \sum_{k\geq 0}(a_k \phi_k(x) + h.c.) - \sum_{k\geq 0}(b_k 
\overline{\phi}_k(x) + h.c.), \ \ \ \lbrack a_k, a^{\dagger}_{k'} \rbrack 
= \delta_{kk'} = -\lbrack b_k, b^{\dagger}_{k'} \rbrack, 
\eeq
and this defines a dS invariant vacuum $\mid \Omega \,\rangle$: 

\beq
a_k \mid \Omega \, \rangle = 0 = b_k \mid \Omega \,\rangle, \ k\geq 
0. \eeq
The whole (Krein-Fock) space $\underline{{\cal H}}$ has a 
Gupta-Bleuler structure which parallels (\ref{GBtri}):
\beq
 \label{QFTGB}
\underline{ \underline{{\cal N}}} \subset \{ 
(a^{\dagger}_g)^{n_0}(a^{\dagger}_{k_1})^{n_1}\cdots 
(a^{\dagger}_{k_l})^{n_l}\mid \Omega \,\rangle \} \equiv 
\underline{{\cal K}} \subset \underline{{\cal H}},
\eeq
where $\underline{ \underline{{\cal N}}}$ is the subspace of the 
physical space $\underline{{\cal K}}$ which is orthogonal to 
$\underline{{\cal K}}$. It is actually the space of gauge states 
since any physical state $\Psi \in \underline{{\cal K}}$  is equal to 
its ``gauge transform'' $\exp{-\frac{\pi\lambda}{H}(a^{\dagger}_g - 
a_g )} \Psi$ up to an element of $\underline{\underline{{\cal N}}}$. 
We shall say that both are physically equivalent. Consistently, an 
observable $A$ is a symmetric operator on $\underline{{\cal H}}$ such 
that $\langle \Psi \mid A
\mid \Psi \, \rangle = \langle \Psi' \mid A \mid \Psi' \, \rangle$ 
for any pair of equivalent physical states. As a matter of fact, the 
field $\varphi$ is \underline{not} an observable whereas 
$\partial_{\mu}\varphi$, where $\mu$ refers to the global coordinates 
(\ref{concoor}), \underline{is}. Therefore the stress tensor 
 \beq
 \label{entens}
T_{\mu \nu} = \partial_{\mu}\varphi \partial_{\nu}\varphi - 
\frac{1}{2} g_{\mu \nu} g^{\rho \sigma} \partial_{\rho}\varphi 
\partial_{\sigma}\varphi \eeq
is an observable. Its most remarkable feature is that it meets all 
reasonable requirements one should expect from such a physical 
quantity, namely, 
\bei
\item No need of renormalization: $\vert \langle k^{n_1}_1 \cdots 
k^{n_l}_l \mid T_{\mu \nu} \mid k^{n_1}_1 \cdots k^{n_l}_l \, \rangle 
\vert < \infty$, 
\item Positiveness of the energy component (\emph{energy} here should be understood in a QFT framework) on the 
physical sector: $\langle k^{n_1}_1 \cdots k^{n_l}_l \mid T_{00} \mid 
k^{n_1}_1 \cdots k^{n_l}_l \, \rangle \geq 0$, 
\item The vacuum 
energy is zero: $\langle \Omega \mid T_{00} \mid \Omega \, \rangle = 
0$.
\ei
The usual approaches to the quantization of the dS massless minimally 
coupled field were precisely plagued by divergences and 
renormalization problems. Here, one can become aware to what extent 
the respect of full de Sitter covariance leads to satisfying physical 
statements, even though the price to pay is to introduce into the 
formalism these (non positive norm) auxiliary states.
\section{Conclusion}
We now arrive at the conclusion of the paper. From its content, we can claim the following. 
\begin{itemize}
  \item In the case of ``massive'' fields, associated with the principal series of the de Sitter group $SO _{\scriptsize 0}(1,4)$, the construction of fields is based on analyticity condition imposed to the Wightman two-point function.
  $${\cal W}_{\nu}(x_1,x_2) = 
\langle \Omega,
\phi(x_1) \phi(x_2) \Omega \rangle,$$ 
  where $\Omega$ is the Fock 
vacuum and $\phi$ is the field operator.
    \item In the case of ``massless'' fields (\emph{e.g.} minimally coupled massless field or conformally coupled fields), 
    associated to the discrete series of  $SO_{\scriptsize 0}(1,4)$, the quantization scheme is  of the 
Gupta-Bleuler-Krein type.
\item The next step  logically consists in the construction of a consistent ``de Sitter QED'', since we now have all elementary bricks (``massive'' spin-$1/2$ field and ``massless'' vector field) to set up 
gauge invariant Lagrangian. But then arises the fundamental question of a measurement guideline/interpretation consistent with dS relativity. A first step should consist in controlling the ``minkowskian'' validity of such a theory through expansion of various quantitative issues of computation in powers of the curvature.
\end{itemize}

In any fashion let us insist on the fact that  relativity principles based on the theory of groups and of their 
representations is one the corner stones of Physics. We hope that the 
present review which deals with de Sitter relativity offers another 
convincing illustration of this well-known (but once too often 
forgotten?) textbook statement.

\end{document}